8-17-2021

# Digital Media and Information Literacy: A way to Paperless Society

Subaveerapandiyan A
*Regional Institute of Education Mysore*, subaveerapandiyan@gmail.com

Anuradha Maurya
*SL Education Institute, Moradabad*, connectanu2net@gmail.com

# Digital Media and Information Literacy: A way to Paperless Society

**Subaveerapandiyan A***

*Professional Assistant*

*Regional Institute of Education Mysore, India*

ORCID: https://orcid.org/0000-0002-2149-9897

**Anuradha Maurya****

*Assistant Professor*

*Dept. of Library and Information Science*

*SL Education Institute, Moradabad, Uttar Pradesh, India*

E-mail: connectanu2net@gmail.com

## ABSTRACT

**Purpose -** The study's main objective was to find out the possibility of a paperless library and society with particular reference to Top 60 Universities from QS world University ranking 2021 and their library professionals. ICT knowledge and skills of these LIS professionals and evaluated their digital literacy skills was another aim of this study.

**Design/methodology/approach -** The researchers used the survey method for this study using a structured questionnaire, distributed through the google form to library professionals of world-famous universities, ranked as top 60 in QS World University Ranking. 206 responses were received. The information collected from the respondents has been analyzed using an Excel sheet and SPSS software.

**Findings -** Most professionals are interested in digital learning and adopting paperless learning to contribute to a paperless society. They go for online ways to answer reference queries of users and work in a refined atmosphere. They are learning from digital resources and have support from online platforms if they suffer. Also, they are actively engaging with the digital environment and promoting it too.

**Paper type:** Research paper

1. Introduction

Information is transferable, verifiable and valuable. The digital world is now filled with uncountable data and information that made it rich and has reached the peak of wealth. Information literacy can change the universe of knowledge and digital society. We are surrounded by Data every minute; however, original or fake can be evaluated with literacy skills, performing the most crucial role. This literacy is thoroughly used in each task to manage all kinds of data and later organize and disseminate it promptly. Information literacy is an inborn right to the digital world, i.e., a fundamental functional quality of everyone's life. Digital skill is pivotal for 21st-century professionals and learners, using which the library world can grow to serve better services to its users.

Information literacy is a generic term and intelligence to observe the need for information, when it is required, how to locate that and use it effectively to get desirable outcomes, i.e., it helps in decision making, the discovery of right information resources, enables problem-solving capability, LSRW (listening, speaking, reading and writing) skills and much more. With technological advancements, everyone, including students and employees, is now benefited from social media, a vast platform where resources are available through learners who face difficulties to search and find the needed information. To differentiate between original and fake content, IL skills are required and helps to evaluate the data.

Information Literacy skills support for lifelong learning where LIS professionals develop their literacy skills by:
      i) Attending and presenting in conferences, seminars, webinars and workshops
      ii) Self-learning and learn while working
      iii) Doing the additional courses
      iv) Attending MOOC courses
      v) Preparing & publishing research papers

**1.2 Meaning and Definition**

1.2.1 Information Literacy: It is an ability to recognize the requirement of information and its location or resources so it can be used effectively and applied as per need.

1.2.2 Digital Literacy: It is a skill that is required to learn and work in an environment that is filled with digital technologies such as social media, gadgets, internet platforms, etc.

1.2.3 Media Literacy: It is an ability to recognize different media platforms available to share information. The media are both digital and print, where print includes newspapers, magazines, etc. however, digital one includes presentations, podcasts, emails, etc.

1.2.4 Metaliteracy: This model empowers and promotes critical thinking among learners and enhances the capacity to collaborate in the digital age. It is a consolidated design that supports the production of knowledge, sharing of expertise in collaborative online communities by modifying existing skill-based methods to information literacy.

1.2.5 Transliteracy: It can read, write and interact on various platforms to write, speak, print media, etc., to digital social networks.

The focused universities are highly acknowledged and renowned. The list of top 60 universities that QS World University Ranking gave, is mentioned below:

| Rank | University | Rank | University |
|---|---|---|---|
| 1 | Massachusetts Institute of Technology (MIT), Cambridge, United States | 31 | The Australian National University, Canberra, Australia |
| 2 | Stanford University, Stanford, United States | 32 | King's College London, London, United Kingdom |
| 3 | Harvard University, Cambridge, United States | 33 | McGill University, Montreal, Canada |
| 4 | California Institute of Technology (Caltech), Pasadena, United States | 34 | Fudan University, Shanghai, China (Mainland) |
| 5 | University of Oxford, Oxford, United Kingdom | 35 | New York University (NYU), New York City, United States |
| 6 | ETH Zurich - Swiss Federal Institute of Technology, Zürich, Switzerland | 36 | University of California, Los Angeles (UCLA), Los Angeles, United States |
| 7 | University of Cambridge, Cambridge, United Kingdom | 37 | Seoul National University, Seoul, South Korea |
| 8 | Imperial College London, London, United Kingdom | 38 | Kyoto University, Kyoto, Japan |
| 9 | University of Chicago, Chicago, United States | 39 | KAIST - Korea Advanced Institute of Science & Technology, Daejeon, South Korea |
| 10 | UCL, London, United Kingdom | 40 | The University of Sydney, Sydney, Australia |
| 11 | National University of Singapore (NUS), Singapore, Singapore | 41 | The University of Melbourne, Parkville, Australia |
| 12 | Princeton University, Princeton, United States | 42 | Duke University, Durham, United States |
| 13 | Nanyang Technological University, Singapore (NTU), Singapore, Singapore | 43 | The Chinese University of Hong Kong (CUHK), Hong Kong, Hong Kong SAR |
| 14 | EPFL, Lausanne, Switzerland | 44 | The University of New South Wales (UNSW Sydney), Sydney, Australia |
| 15 | Tsinghua University, Beijing, China (Mainland) | 45 | University of British Columbia, Vancouver, Canada |
| 16 | University of Pennsylvania, Philadelphia, United States | 46 | The University of Queensland, Brisbane, Australia |

| 17 | Yale University, New Haven, United States | 47 | Shanghai Jiao Tong University, Shanghai, China (Mainland) |
| --- | --- | --- | --- |
| 18 | Cornell University, Ithaca, United States | 48 | City University of Hong Kong, Hong Kong, Hong Kong SAR |
| 19 | Columbia University, New York City, United States | 49 | The London School of Economics and Political Science (LSE), London, United Kingdom |
| 20 | The University of Edinburgh, Edinburgh, United Kingdom | 50 | Technical University of Munich, Munich, Germany |
| 21 | University of Michigan-Ann Arbor, Ann Arbor, United States | 51 | Carnegie Mellon University, Pittsburgh, United States |
| 22 | The University of Hong Kong, Hong Kong, Hong Kong SAR | 52 | Université PSL, France |
| 23 | Peking University, Beijing, China (Mainland) | 53 | Zhejiang University, Hangzhou, China (Mainland) |
| 24 | The University of Tokyo, Tokyo, Japan | 54 | University of California, San Diego (UCSD), San Diego, United States |
| 25 | Johns Hopkins University, Baltimore, United States | 55 | Monash University, Melbourne, Australia |
| 26 | University of Toronto, Toronto, Canada | 56 | Tokyo Institute of Technology (Tokyo Tech), Tokyo, Japan |
| 27 | The Hong Kong University of Science and Technology, Hong Kong, Hong Kong SAR | 57 | Delft University of Technology, Delft, Netherlands |
| 28 | The University of Manchester, Manchester, United Kingdom | 58 | University of Bristol, Bristol, United Kingdom |
| 29 | Northwestern University, Evanston, United States | 59 | Universiti Malaya (UM), Kuala Lumpur, Malaysia |
| 30 | University of California, Berkeley (UCB), Berkeley, United States | 60 | Brown University, Providence, United States |

## 2. Review of Literature

Demirel and Akkoyunlu (2017) used a relational descriptive model of participants from elementary education teachers of Turkey. The study's main aim is to know whether information literacy is used for lifelong learning or not. The result shows that schools are promoting information literacy for lifelong learning; teachers have a strong propensity for lifelong learning with literacy skills. It also emphasized that lifelong learning is possible in the digital society with digital literacy skills as well as depends upon a student's strong will. A trained teacher community can teach the students well.

Gretter and Yadav (2018) identified pre-service teacher's perceptions about teaching media literacy skills. The study was conducted with semi-structured interview questions. The reason for conducting this study is that they found the gap between the teachers learning media literacy skills; endless benefits possible with media literacy, but the teachers lack knowledge. Instead, they assume ML skill is essential to the students but did not emphasize it in the teacher education curriculum. It also found that students have a favorable opinion about media literacy, but they expect teachers to add media literacy as a part of the subject in class. Also, it reveals pre-service teachers regularly used technology and social media apps for up-to-date news and information even though they are comfortable using digital media. However, they need the training to know how to teach media information literacy in the classroom.

Yevelson-Shorsher and Bronstein (2018) examined students, librarians and faculties information literacy and their perspectives. The study was conducted with semi-structured interviews and revealed that students agree about having less knowledge on information literacy as they did not get adequate help from their faculties; also, students are in-cognizant of the library resources and services offered by the libraries. So, they required that librarian's need to promote and market the library services to the user community. The study emphasizes that faculties and librarians have to teach or conduct training classes to support the students with information literacy.

Botturi (2019) conducted a case study on a two-credit introductory course over Digital and Media Literacy Education in Switzerland, where teachers and students were targeted for investigation. The pre/post survey was done to collect data and interviews to get a clear picture of the pre-service approach of teachers to DML. The study revealed that despite scarce and limited resources, it affected students and enabled faculties to blend media education and digital literacy domains in their profession.

Sharun (2019) conducted a study using semi-structured interviews to present a detailed exploration of the ACRL Framework (Association of College and Research Library) by implementing it in the professional workflow of health and human service working at a community health centre. This was to define the way professionals experience information literacy in their workspace. They observed the experience of professional workers to know the importance and nature of the information they use in their personal and professional lives, which is dedicated to their work atmosphere.

Subekti et al. (2019) conducted a descriptive and survey study on the information literacy skills of secondary school teachers from Indonesia and focused on teacher's scientific writing skills. They mainly discussed copyright issues, information resources, and scientific writing. Only a few teachers were excellent at scientific writing, around 31% of total but slightly familiar with paraphrasing. Hence, they need more training and practice on information literacy. They found teachers are seeking information from online sources instead of print ones.

Julien et al. (2020) studied community college librarians of Florida and New York using an online survey to extract information literacy needs, strengths and weaknesses in them along with students. They showed the influence of the ACRL Framework (Association of College and Research Library) on librarians, challenges they face, and success in implementing their work. The collected data shows the fortune to support and enhance instructions to create future professionals more successful in their work.

Lebid and Shevchenko (2020) studied the critical measurement used to develop critical thinking skills. They used TRIZ (theory of inventive education) and ARPS (algorithms for solving problems situations) in media education. The capacity of strategic planning, problem-solving ability using creativity can be cultivated. Case problems, system mapping, and system thinking tests can be done using skills and tools to be developed using media literacy and its education, which will transform traditional education methods and processes.

Wade et al. (2020) conducted a study in English schools of Canada to know about strategies of web-based inquiries of Information Society of Twenty-First Century (ISIS-21) and its impact on elementary students by improving information literacy skills. ISIS-21 used multimedia series and principles as its foundation and development. A trial was held focusing 150 students in two phases where research design was one of them, and the other was data collection and teacher self-reports. This study established the usefulness and importance of using ISIS-21 at school that could promote the growth of IL skills among students.

Eger et al. (2021) researched public universities of the Czech Republic, Slovakia and Poland. They focused on communication activities on social media network Facebook and collected quantitative data from 24 profiles, which was analyzed using Netvizz tool. The analyzed content was used to examine the influence. The result varies from university to university on Facebook usage and supports a combined view on content marketing. The communication strategy of universities using Facebook or other social networking sites is disadvantageous to its mission of dissemination of results of research.

Jamshed, Jamshaid and Saleemi (2021) determined to study using a survey to know the pattern of library usage by students of Law studies of public universities of Punjab, Pakistan to analyze the information requirement of law students, the purpose of library visits, and services provided by libraries along with the issues faced by students of law in their libraries. The collected data was analyzed by SPSS V23 that contained frequencies, mean, standard deviation, percentage and mode. The findings revealed that students visited the library once a month for assignments and preparation for exams. The main problem was discovered that online resources were absent, legal research journals. Also, no law librarian was there, and all this was recommended to improve that would help legal education in Pakistan.

Lamont (2021) investigated the view about information literacy and how it is helpful for the uninterrupted growth in professional and workplace from secondary teachers of Scottish schools. She used a qualitative and semi-structured interview method for collecting the data and carried out how secondary teachers developed their literacy skills, analysis based on their demographic backgrounds such as educational qualifications and personal development. The teacher's understanding of information literacy was performed a textual analysis based on the Scottish syllabus. The result discovered the teacher's lack of familiarity in information literacy which was not given the workplace development and career developments. The author suggested that teachers be more aware of literacy skills, misinformation and evaluation of information sources.

Zakharov & et al. (2021) conducted a survey study on the digital world and digital literacy competencies faced by teachers and found that 21st century teachers were strong in technical devices used and related domain knowledge but were not much aware of how to combine it with teaching pedagogical activities. The result revealed 50% of teachers' average ICT skills, 22.6% of teachers only prepared to adopt digital resources.

3. **Research Method**

The study used mixed research methods of library professionals and discussed their digital information literacy and the possibilities of a paperless society. For this study, we conducted a user survey focusing on the top 60 universities from QS World University Ranking 2021, determined to know about perspective and practices of using digital media and knowledge of information literacy, paperless society. A structured questionnaire prepared, distributed through E-mails among library professionals of the top 60 universities of the world as per QS World University Ranking. The questionnaire was divided into two sections, first containing demographic information of participants and other parts had questions related to type of records, digitization of library, computerization of library, awareness of professionals regarding different resources along with questions on information literacy, digital media, paperless learning, etc. They were asked about problems dealing with digital content, resources and views on digital and information literacy. Few questions were designed using the Likert Scale, and responses were analyzed with statistics defined in percentage, mean, standard deviation, and tabulated and graphical presentation.

4. **Data Analysis**

The findings of the survey study are mentioned in tabular and graphical form. All responses are counted and defined in percentage, whereas few queries are mentioned using mean and standard deviation.

| Table 1. Demographic details of respondents | | | |
|---|---|---|---|
| **Demographic profile** | **Parameters** | **Frequency** | **Percentage** |
| Age | 18-24 | 0 | 0.0 |
| | 25-30 | 18 | 8.7 |
| | 31-40 | 73 | 35.5 |
| | 41-50 | 69 | 33.5 |
| | 51 and above | 46 | 22.3 |
| Gender | Male | 61 | 29.6 |
| | Female | 145 | 70.4 |

| Working Experience | < 1 year | 0 | 0.0 |
| --- | --- | --- | --- |
| | 1-5 years | 25 | 12.1 |
| | 6-10 years | 51 | 24.7 |
| | 11-20 year's | 68 | 33.1 |
| | Over 20 years | 62 | 30.1 |
| **Total** | | **206** | **100** |

Above table 1 describes the demographic details of the respondents. The respondents belong to almost all age groups from 25 years to 51+ years, where the highest respondents(age-wise) are of group 31 to 40 years of age which is 35.5%, followed by 41-50 years of age making 33.5%, whereas no respondents are of age group 18 to 24 years. Gender wise, the highest number of respondents are females, 70.4% though males are only 29.6%. This result clearly shows that more women work in the libraries than men, performing as passive respondents. To work experience, significant respondents have earned 11-20 years of experience making it 33.1%, followed by professionals having 20 years of experience, 30.1%; however, no respondent found with less than one year of experience. So, above the table clarifies this study is conducted with highly experienced professionals working in libraries.

**Figure 1. The record maintains in the library?**

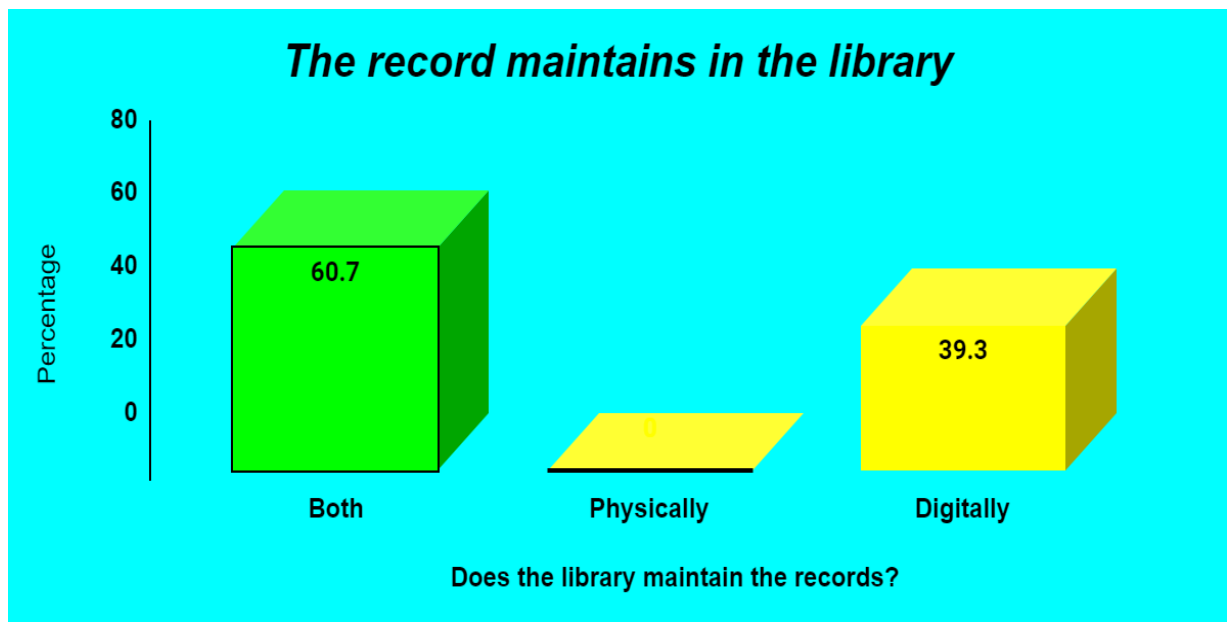

It is clear from the above figure 1, defining the maintenance of the library records, libraries maintain it digitally and physically by 60.7%. Only 39.3% keep it safe digitally, whereas no libraries are maintaining it physically. It clarifies that libraries of developed countries have almost migrated from physical to digital form, so it's possible to adopt the records in the digital form entirely in upcoming years.

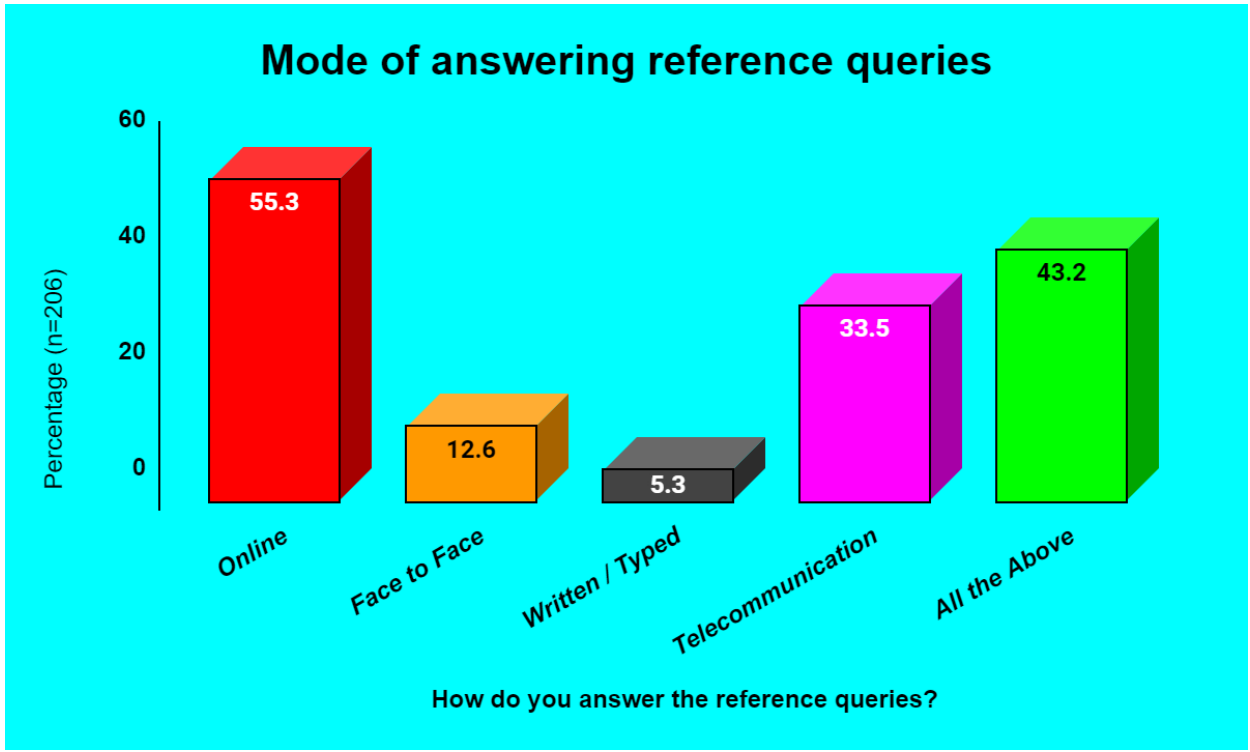

**Figure 2. Mode of answering reference queries**

As mentioned in figure 2, 55.3% of library professionals answer the reference queries online, 12.6% prefer face-to-face communication, 5.3% share it by writing/in typed form. 33.5% of them use telecommunication, and 43.2% of professionals use all of them to solve queries. So, the minor questions are answered in written/typed form, which is only 5.3% of the total though the highest queries answered online with 55.3%. Hence, professionals are adopting the online mode of services most.

| Table 2. Library digitization percentage | | |
|---|---|---|
| **Percentage of library digitized** | **Respondents** | **Percentage** |
| Not digitized | 3 | 1.4 |
| Below 30% | 50 | 24.3 |
| 31-50% | 70 | 34.0 |
| 51-70% | 40 | 19.4 |
| 71-90% | 33 | 16.0 |
| Fully digitized | 10 | 4.9 |

**Figure 3. Library digitization percentage**

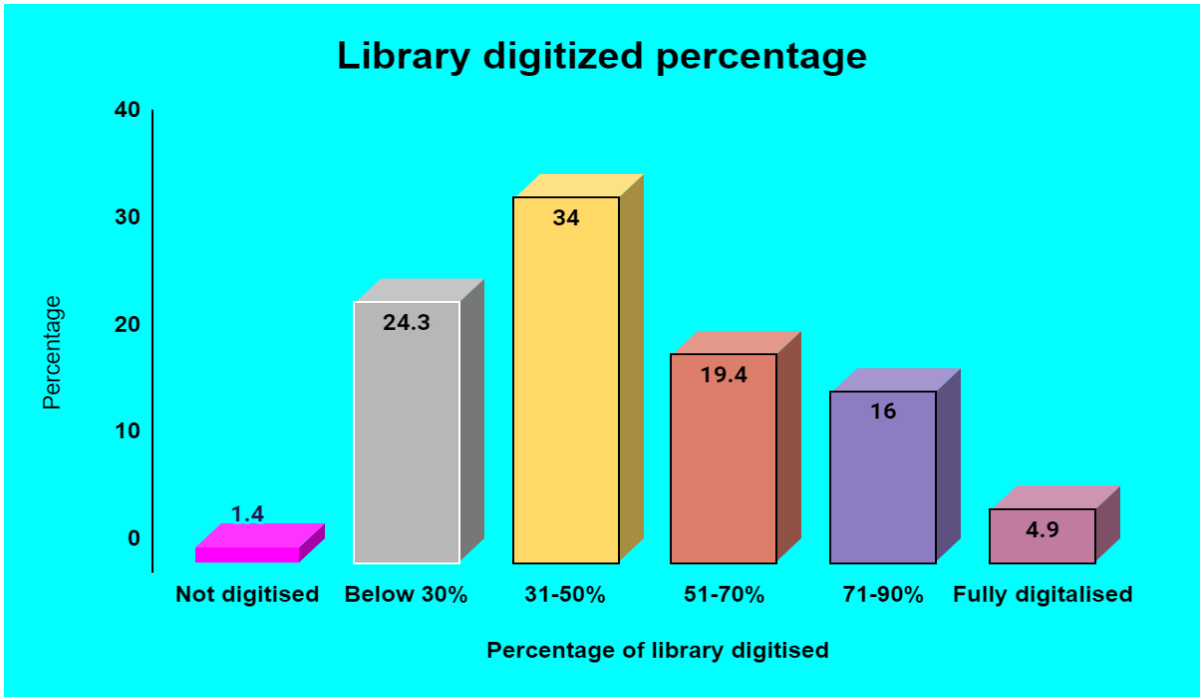

According to the results of above table 2 and figure 3, the highest percentage of the digitized libraries is 31 to 50% as per responses. In contrast, only 1.4% of libraries are not-digitized, but 4.9% of libraries have received complete digitization. This again shows that libraries are moving towards digitization and are under process.

**Figure 4. Computerization of library status**

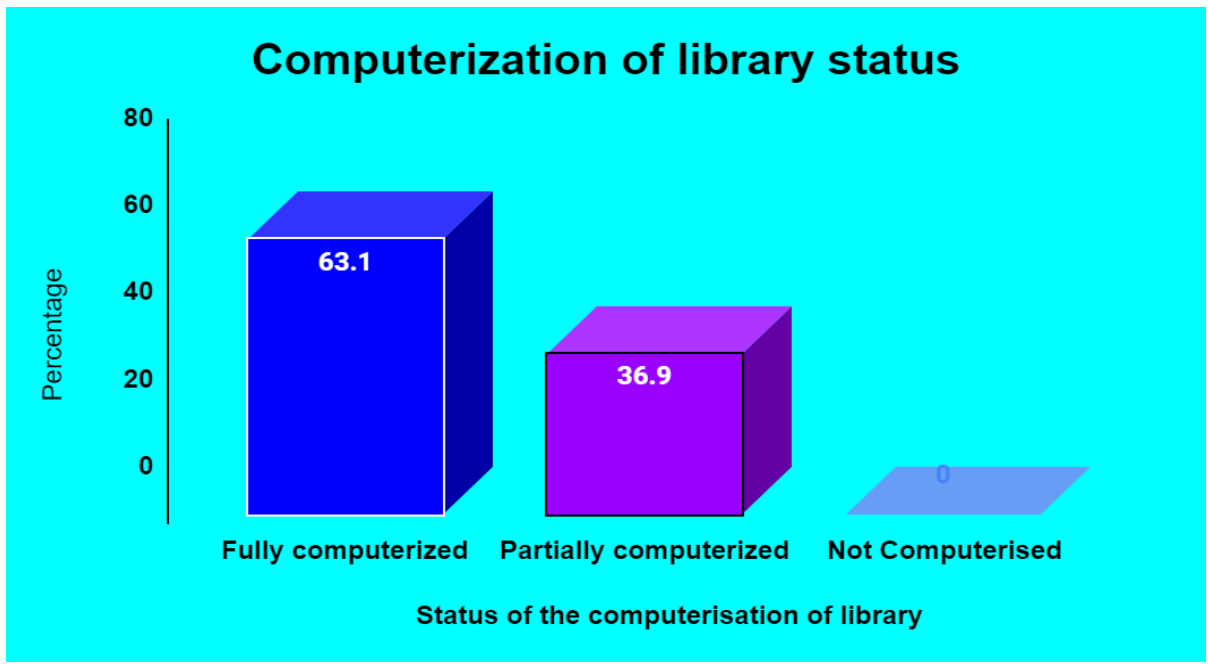

From the results of the above figure 4, the status of the computerization of the libraries is well satisfied with 63.1% of them fully computerized and only 36.9% are partially computerized. It also shows that no library is left behind in the race of computerization.

**Figure 5. Way of collecting library fine**

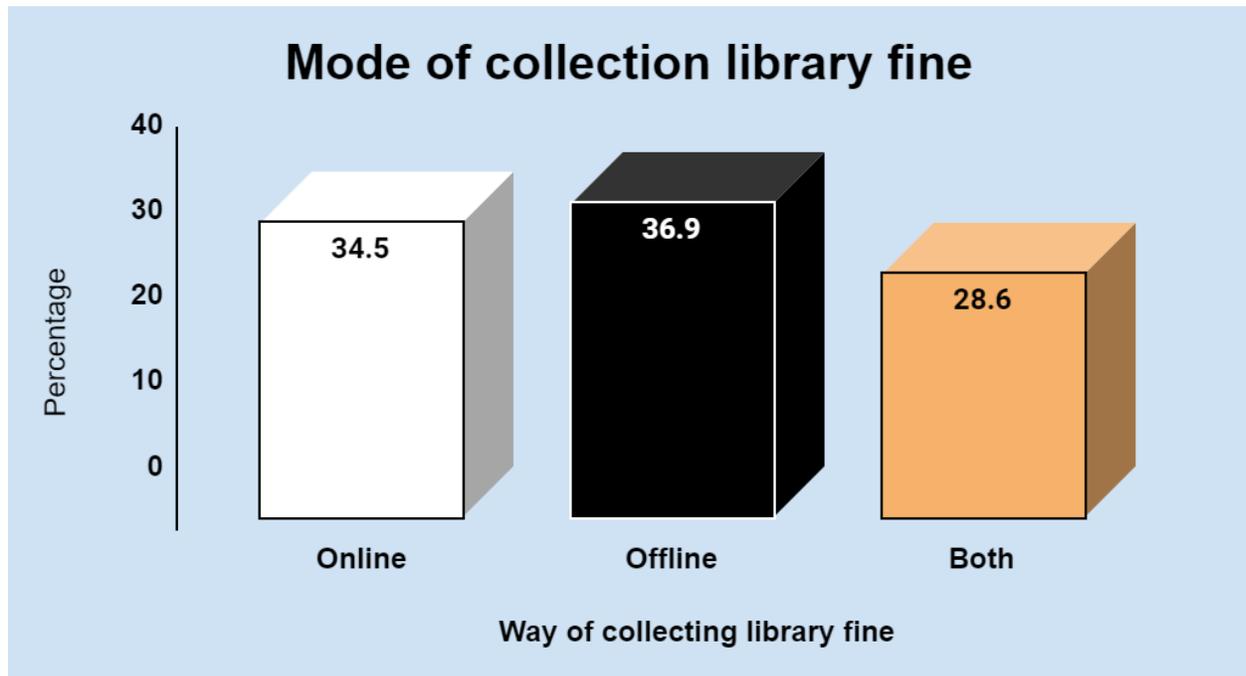

Figure 5 shows that libraries follow both ways of collecting library fines, i.e., online/offline, 28.6%. Libraries that receive fine offline are only 36.9% through online fine collection, which is preferred by 34.5% of libraries. It is clear evidence that paperless transactions in the libraries are more favored than the offline mode of collection of fines. Yet, many of them use both online and offline ways of collecting library fines.

| Table 3. Awareness of open sources and software | | |
|---|---|---|
| **Awareness** | **Yes** | **No** |
| OER | 192 (93.2%) | 14 (6.8%) |
| MOOC | 199 (96.6%) | 7 (3.4%) |
| Creative Commons | 195 (94.7%) | 11 (5.3%) |
| Open Source Software | 206 (100%) | 0 (0.0%) |
| Institutional Repositories | 206 (100%) | 0 (0.0%) |

**Figure 6. Awareness of open sources and software**

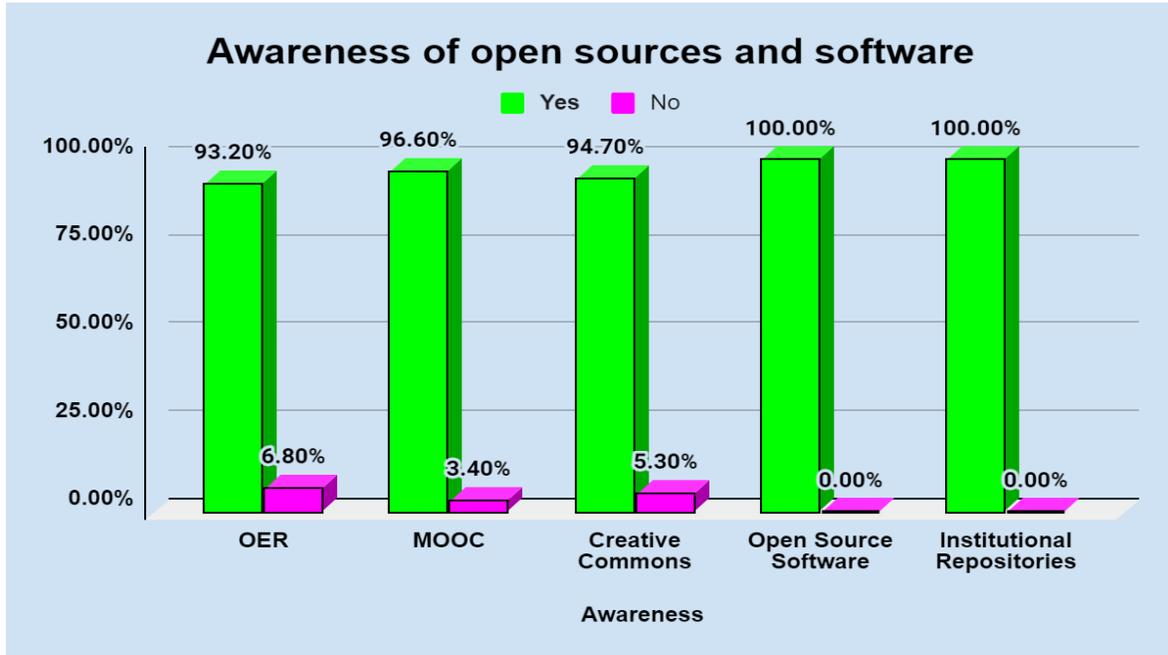

Table 3 and figure 6 showed awareness of academic literacy, 93.2% uses Open Educational Resources, Massive Open Online Course used by 96.6%, Creative Commons used by 94.7% however Open Source Software and Institutional Repositories used by all with 100% of respondents. It shows us, librarians are good at digital literacy and have an awareness of open-source software.

**Figure 7. MOOC course registered**

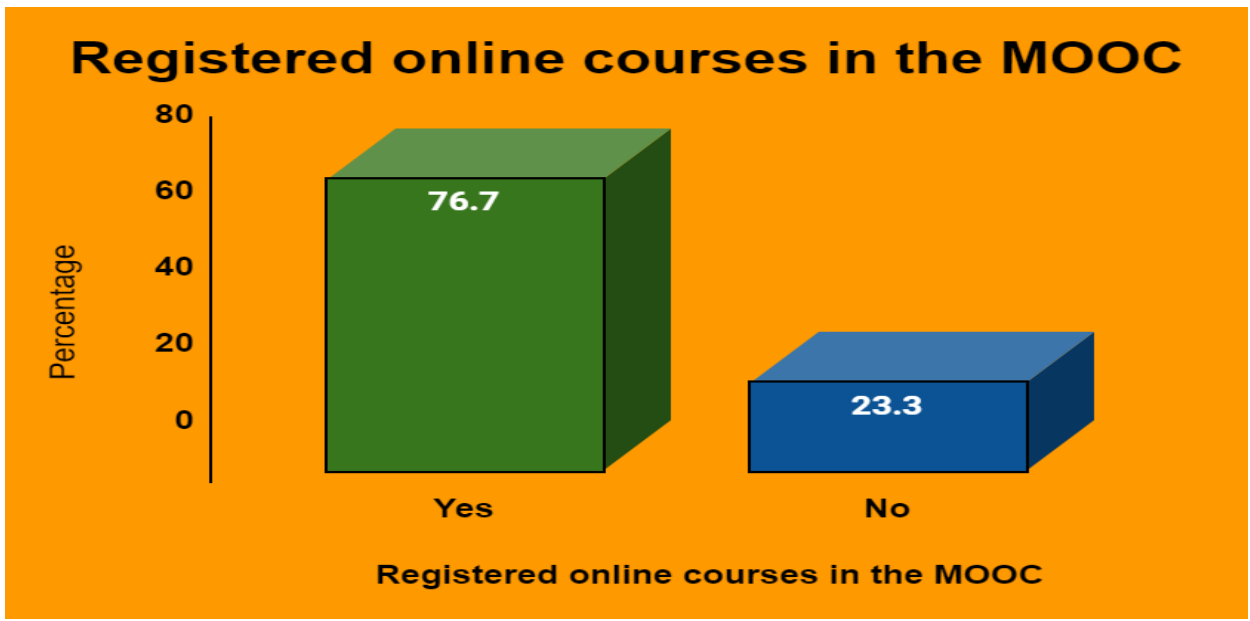

As stated in figure 7, which is about registration in online courses on the MOOC platform. 76.7% of respondents are successfully registered, whereas 23.3% are still away from it. The result shows us most of them migrated towards online courses and positively following.

**Figure 8. Online reading habit**

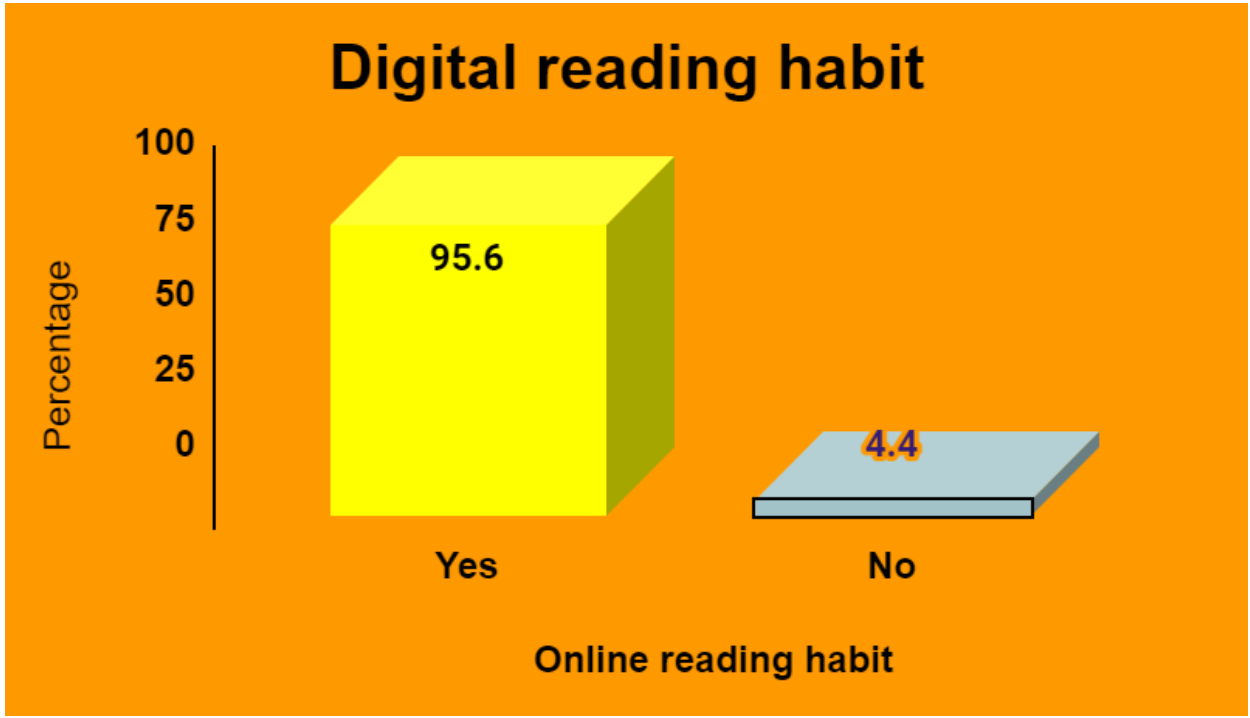

Figure 8 reveals that the digital/online tools used for the reading habit by library professionals are in the majority with *'yes'* 95.6% and very few of them mentioned not having a habit of digital reading, which is 4.4%. It shows that professionals prefer to read online rather than print.

| Table 4. Opinion on online/digital Reading | | |
|---|---|---|
| **Opinion on online/digital reading** | **Respondents** | **Percentage** |
| Comfortable | 114 | 55.3 |
| Uncomfortable | 27 | 13.1 |
| Both | 65 | 31.6 |
| Total | 206 | 100 |

**Figure 9. Opinion on online/digital reading**

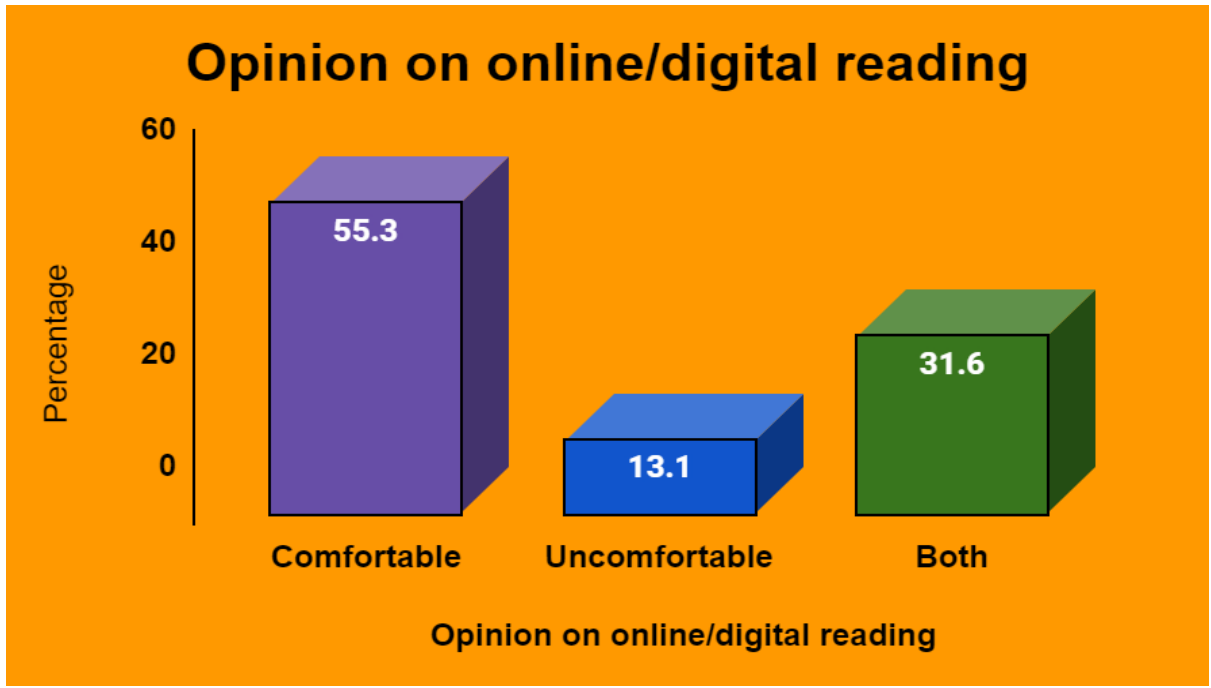

The data is displayed in above table 4 and figure 9 over opinion on online/digital reading. The majority of the respondents i.e., 55.3%, are comfortable with online reading. The other 31.6% said they are both comfortable and uncomfortable as per content and comfort. Rather, very few told about their discomfort with online reading, which is 13.1%.

**Figure 10. Difficulties in vocabulary, you would search by**

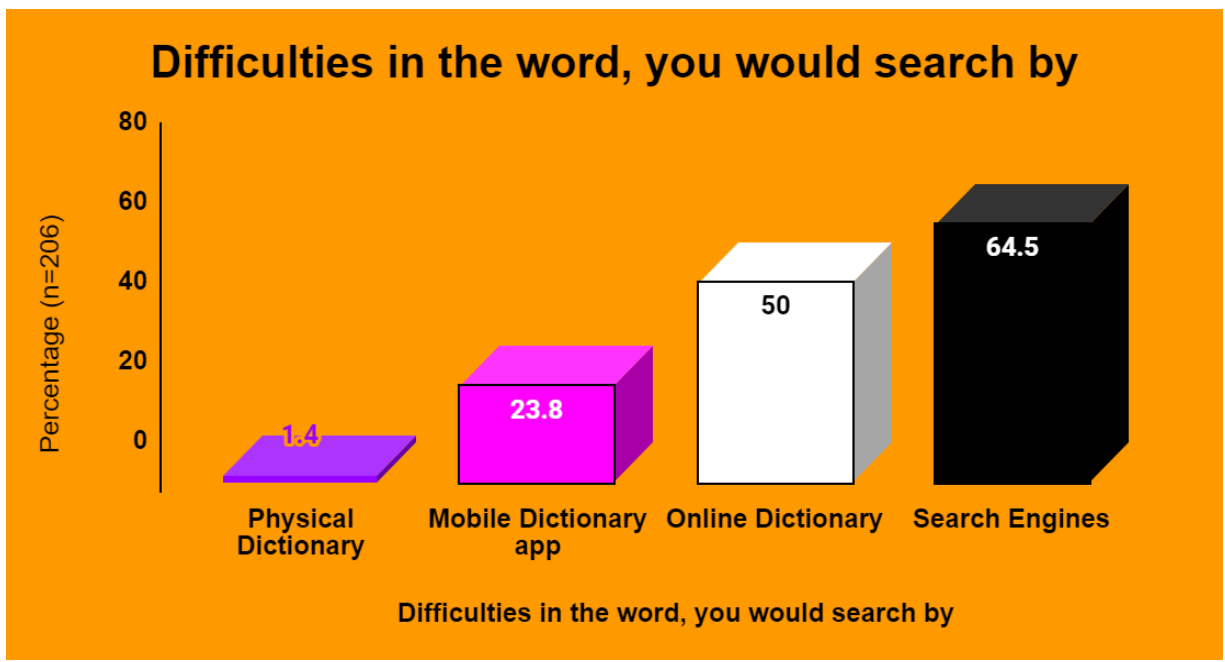

The above figure 10 shows whenever they face difficulty over vocabulary; professionals prefer using search engines to find/search about difficult terms found 64.5% out of all, followed by an online dictionary with 50% and other 23.8% choose mobile applications however very less use physical dictionary which is 1.4%. The above table shows that the 21st generation is migrating from physical to digital forms of dictionaries and search engines to solve related vocabulary issues.

**Figure 11. Social media using**

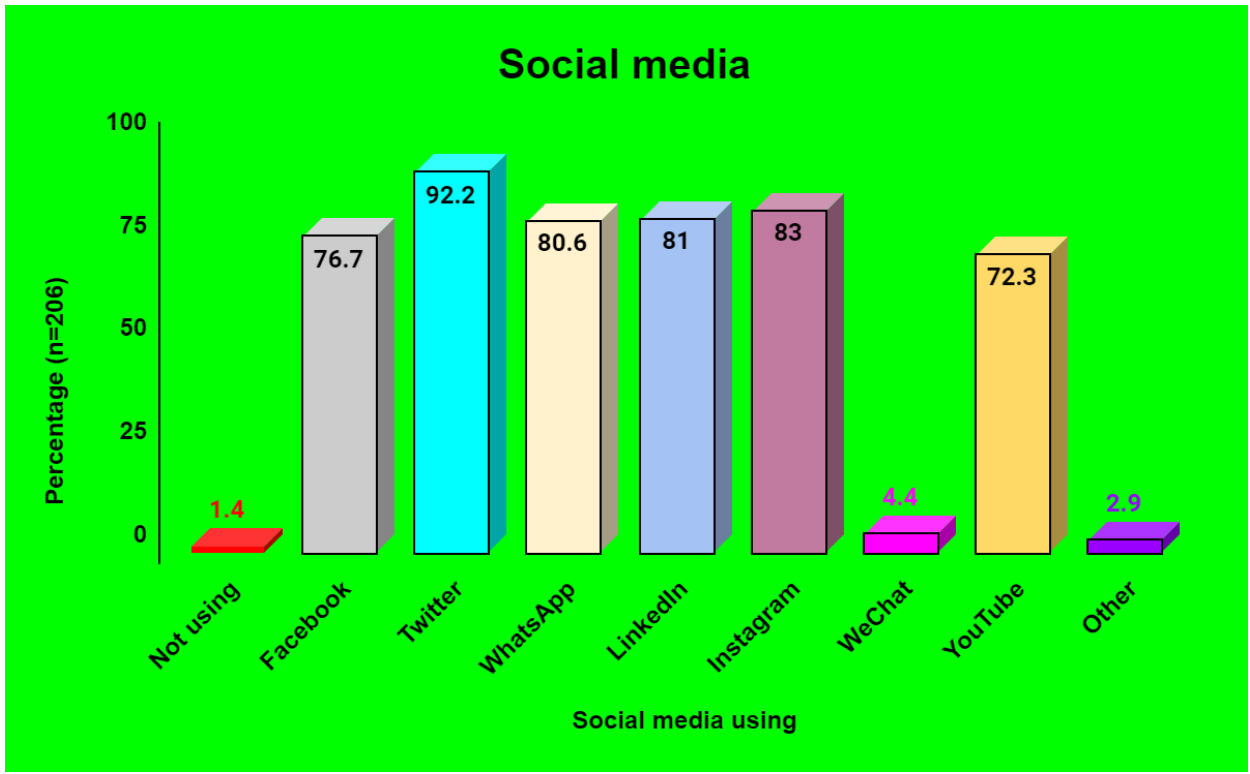

Figure 11 shows social media applications used by library professionals. 92.2% that makes a majority of the respondents uses Twitter, followed by Instagram users with 83% and LinkedIn with 81%, whereas WhatsApp used by 80.6% of professionals and 76.7% use Facebook. 72.3% use YouTube and 4.4% of them prefer WeChat. Only 2.9% of professionals use other applications, and 1.4% do not use social media apps. It defines an inclination towards Twitter and Instagram and LinkedIn and WhatsApp though the rest have mixed preferences.

| Table 5. Internet and cybersecurity awareness | | |
|---|---|---|
| Awareness | Yes | No |
| Internet Security | 206 (100%) | 0 (0.0%) |
| Cybersecurity | 206 (100%) | 0 (0.0%) |

Table 5 displays that all widely adept Internet & Cybersecurity awareness. 100% of library professionals are aware of cyber and Internet security, which is an excellent sign of moving towards technology.

| Table 6. Accessing of paperless resources | | |
|---|---|---|
| **Opinions** | **Opinion of the paperless society** | **Opinion of paperless library** |
| 0% | 16 (7.8%) | 16 (7.8%) |
| 10 to 30% | 41 (19.9%) | 44 (21.4%) |
| 40 to 60% | 54 26.2%) | 58 (28.1%) |
| 70 to 90% | 68 (33%) | 82 (39.8%) |
| 100% | 27 (13.1%) | 6 (2.9%) |

**Figure 12. Accessing of paperless resources**

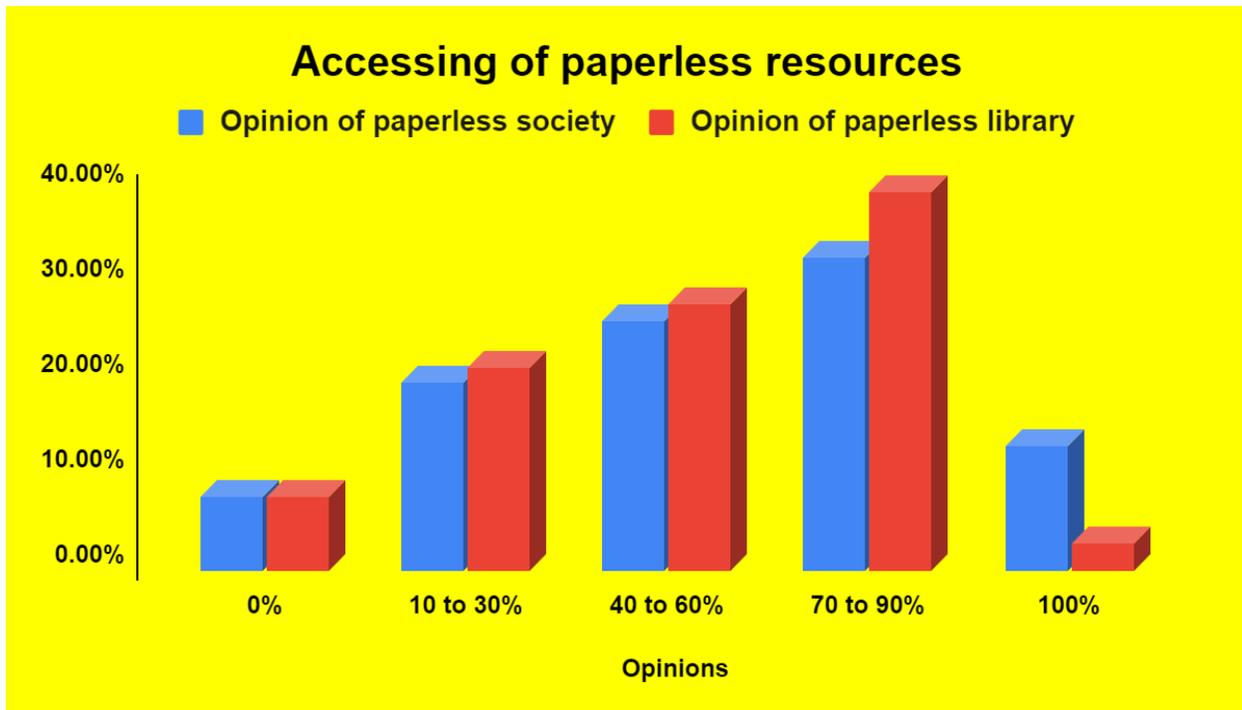

A study of data in table 6 and figure 12 shows that 70 to 90% opinion shares view on awareness in paperless content/resources usage with 39.8% and 33% of all following paperless society. 7.8% have no opinion over paperless society and resources.

**Figure 13. Digital Media and Information Literacy are one of the ways to a paperless society**

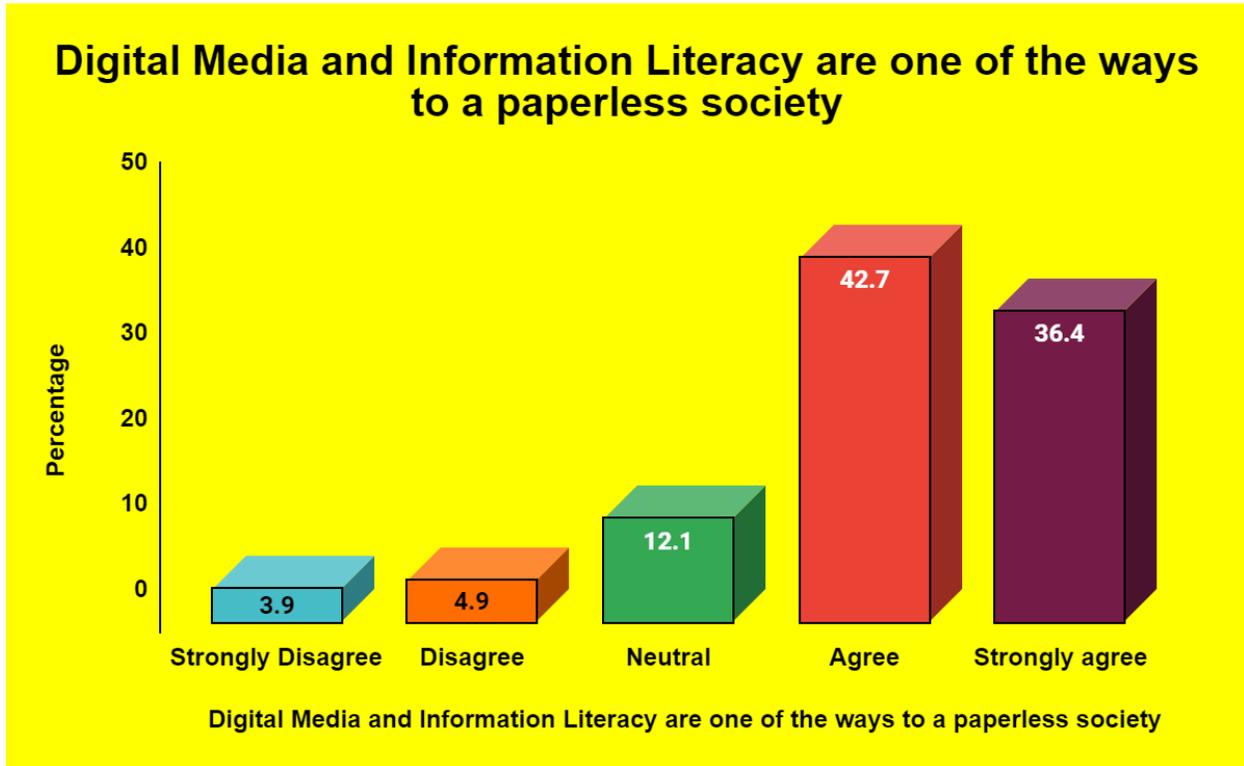

As shown in figure 13, DMIL is one of the ways to paperless society as per 42.7% agreement, and 36.4% agreed strongly; hence combined 79.1% says it's possible. With literacy, we can change the view of 3.9% of professionals as it is the ultimate tool for all kinds of digital communication.

| Table 7. Paperless Learning |||
|---|---|---|
| **Paperless learning environment** | **Mean** | **Standard Deviation** |
| Digital Classroom | 4.07 | 0.98 |
| Electronic Thesis and Dissertation (ETD) | 4.2 | 0.93 |
| E-Resources | 4.44 | 0.77 |
| Online Learning | 4.14 | 0.82 |
| OER | 4.3 | 0.82 |
| Search Engines and their Usage | 4.17 | 0.81 |
| Smart Classroom | 3.93 | 1.01 |
| Virtual Classroom | 3.99 | 0.96 |
| Library Automation | 4.06 | 0.91 |

**Figure 14. Paperless learning environment**

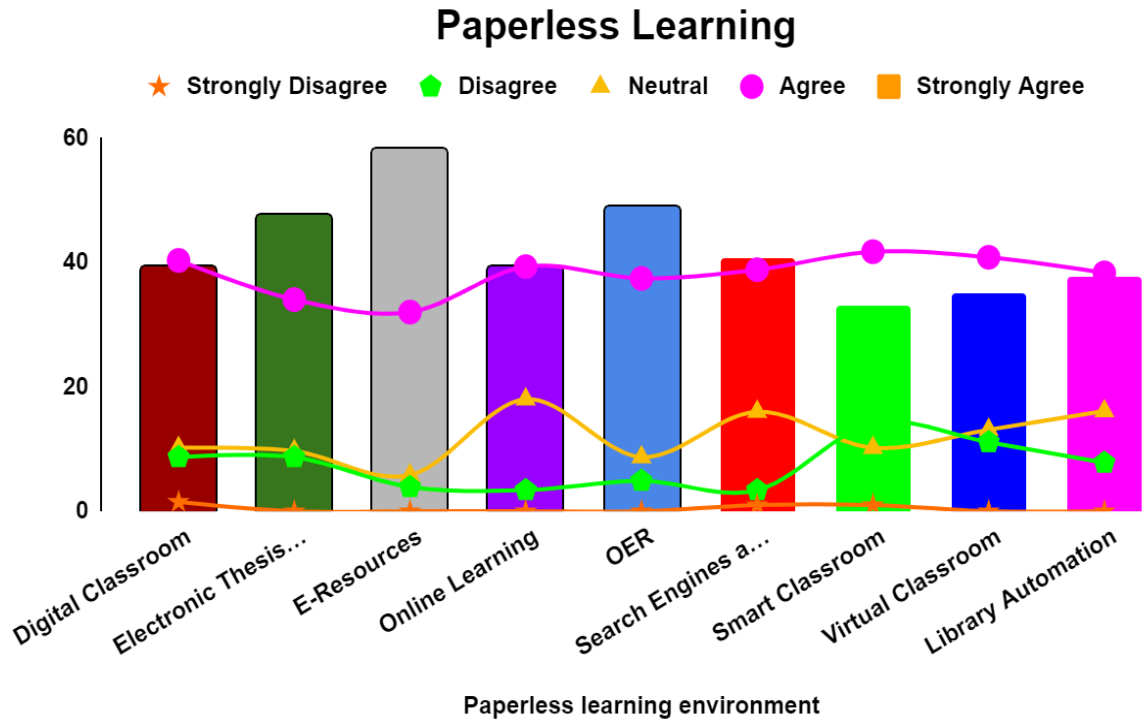

Table 7 and figure 14 explains about inclination towards paperless learning where all the statements (8) are near mean value 4 which means E-resources (M=4.44, SD=0.77) and OER (M=4.30, SD=0.82), ETD (M=4.20, SD=0.93), search engine and usage (M=4.17, SD= 0.81), online learning (M=4.14, SD=0.82), digital classroom (M=4.07, SD=0.98), library automation (M=4.06, SD=0.91), virtual classroom (M=3.99, SD=0.96), smart classroom (M=3.93, SD=1.01) are platforms where respondents are inclined and agrees to take part for paperless learning.

| Table 8. Digital Literacy will increase learning ability | | |
|---|---|---|
| **Digital Literacy will increase learning ability** | **Mean** | **SD** |
| RFID | 3.5 | 1.04 |
| Smartcard | 3.54 | 0.98 |
| Barcode | 3.57 | 0.95 |
| Artificial Intelligence | 3.89 | 0.89 |
| Web 2.0 and 3.0 | 3.72 | 0.87 |
| IoT | 4.3 | 0.81 |
| Cloud computing | 4.31 | 0.71 |

Scale. 1=Strongly disagree. 2=Disagree. 3=Neutral. 4=Agree. 5=Strongly Agree *SD= Standard Deviation*

**Figure 15. Digital Literacy will increase learning ability**

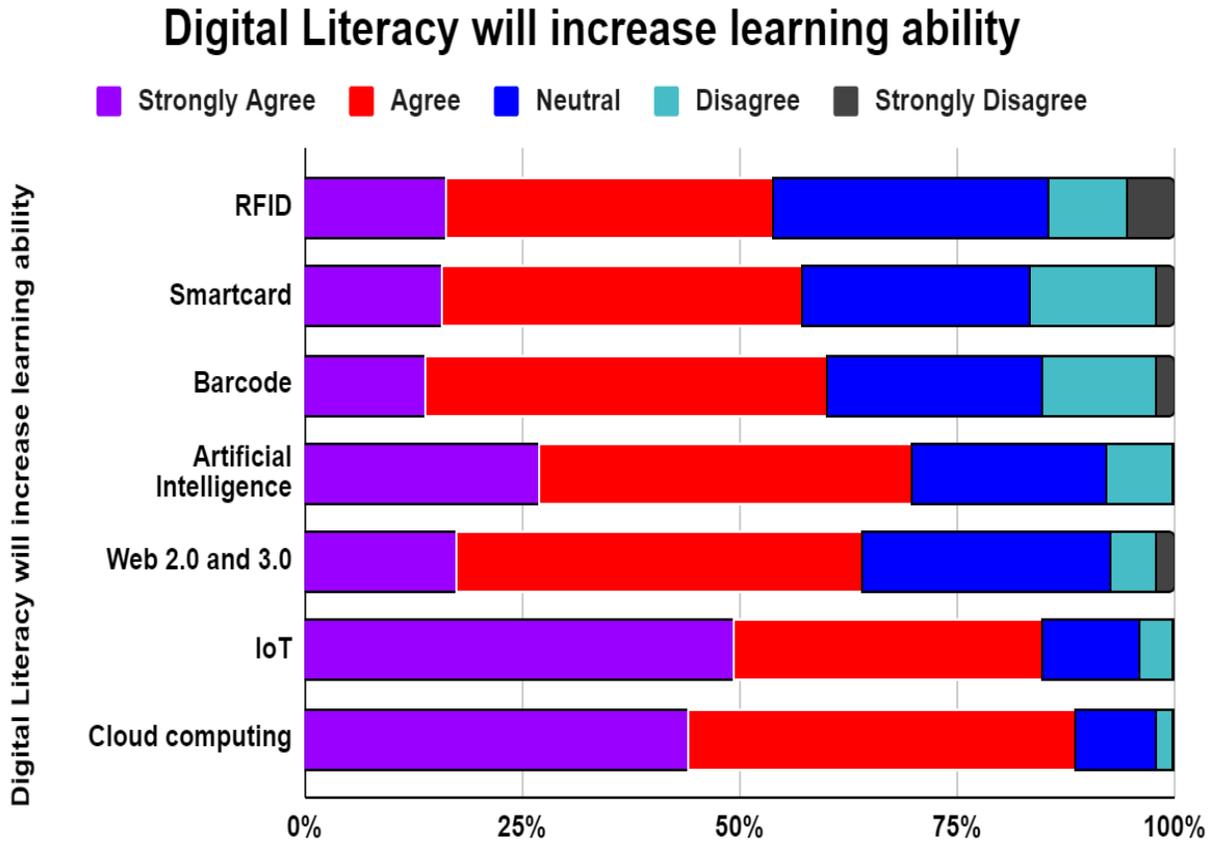

Table 8 and figure 15 shows that digital literacy will promote learning ability where 4 ways mean values are near 4 i.e., cloud computing (M=4.31, SD=0.71), IoT (M=4.30, SD=0.81), artificial intelligence (M=3.89, SD=0.89), web 2.0 & 3.0 (M=3.72, SD=0.87), that shows the agreement of respondents over learning ability will get promoted by digital learning. 2 ways are near mean value 3, RFID (M=3.5, SD=1.04) and Smart card (M=3.54, SD=0.98) showing their neutral behavior towards digital learning.

| Table 9. Awareness of ICT | | | | | |
|---|---|---|---|---|---|
| **Awareness of ICT** | **Excellent (%)** | **Above Average (%)** | **Average (%)** | **Below Average (%)** | **Extremely Poor (%)** |
| OS Linux | 28.6 | 16.1 | 21.8 | 24.8 | 8.7 |
| MS Office | 40.3 | 46.1 | 12.1 | 1.5 | 0 |
| Web page design | 21.4 | 32 | 31.6 | 8.7 | 6.3 |
| Photoshop | 19.9 | 32 | 28.7 | 14.1 | 5.3 |
| Create metadata | 35.9 | 39.3 | 14.6 | 8.7 | 1.5 |
| Customization of software | 17.5 | 32.5 | 19.9 | 18.4 | 11.7 |
| Database Management | 18.4 | 33 | 23.8 | 18 | 6.8 |
| RFID Technology | 24.3 | 32 | 25.7 | 14.6 | 3.4 |
| Bibliometric Software | 25.7 | 32 | 24.3 | 10.7 | 7.3 |
| SPSS | 12.6 | 21.8 | 19 | 22.4 | 24.2 |

Figure 16. Awareness of ICT

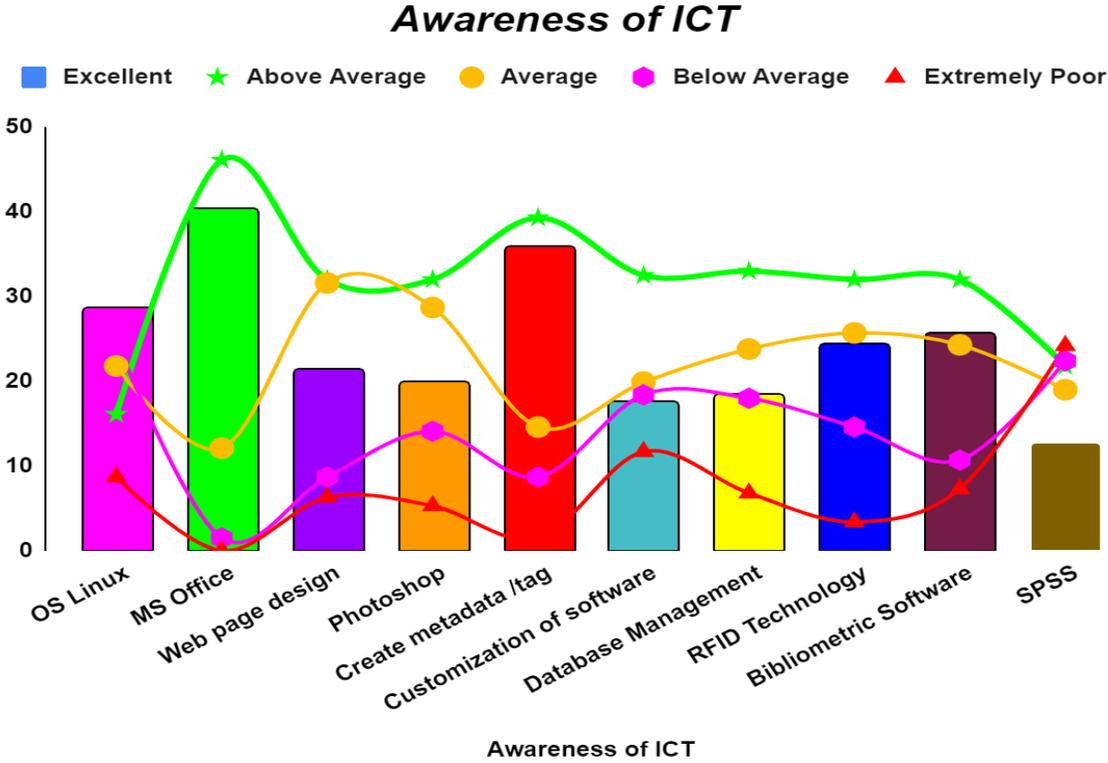

Table 9 and figure 16 shares details about awareness of ICT. It shows that most of them are aware of the latest technologies used by libraries and information services with above-average knowledge of these tools/technologies. Yet, professionals need to focus more on existing, upcoming and emerging technologies in the field.

**Conclusion**

This study is beneficial for library professionals of focused universities because the findings show that most library professionals who took part in the survey are mature and qualified professionals and are positively inclined towards digital literacy and the use of social media platforms. The majority of them are interested in digital learning, but few are there who want to continue with traditional ways. They are adopting paperless learning and contributing towards a paperless society. Professionals prefer online ways to serve reference queries of users and work in a computerized and digitized atmosphere. They are eagerly learning from digital resources and seeking help from digital platforms. Also, they are actively participating in the digital environment and promoting it too. It is recommended that those who have less interest or are less familiar with technology must work on their digital literacy skills to become more competitive with other professionals and could inspire upcoming professionals in the field.